\documentclass[fleqn,10pt]{wlscirep}
\usepackage[utf8]{inputenc}
\usepackage[T1]{fontenc}
\usepackage{caption}
\usepackage{graphicx}
\usepackage[version=4]{mhchem}

\title{Lattice Thermal Transport in Two-Dimensional Alloys and Fractal Heterostructures}

\author[1]{Aravind Krishnamoorthy}
\author[1]{Nitish Baradwaj}
\author[1]{Aiichiro Nakano}
\author[1]{Rajiv K. Kalia}
\author[1,*]{Priya Vashishta}
\affil[1]{Collaboratory for Advanced Computing and Simulations, University of Southern California, Los Angeles, CA 90089}
\affil[*]{priyav@usc.edu}


\begin{abstract}
  Engineering thermal transport in two dimensional materials, alloys and heterostructures is critical for the design of next-generation flexible optoelectronic and energy harvesting devices. Direct experimental characterization of lattice thermal conductivity in these ultra-thin systems is challenging and the impact of dopant atoms and hetero-phase interfaces, introduced unintentionally during synthesis or as part of deliberate material design, on thermal transport properties is not understood. Here, we use non-equilibrium molecular dynamics simulations to calculate lattice thermal conductivity of \ce{(Mo|W)Se_2} monolayer crystals including \ce{Mo_{1-x}W_xSe_2} alloys with substitutional point defects, periodic \ce{MoSe_2}$|$\ce{WSe_2} heterostructures with characteristic length scales and scale-free fractal \ce{MoSe_2}$|$\ce{WSe_2} heterostructures. Each of these features has a distinct effect on phonon propagation in the crystal, which can be used to design fractal and periodic alloy structures with highly tunable thermal conductivities. This control over lattice thermal conductivity will enable applications ranging from thermal barriers to thermoelectrics.
\end{abstract}

\begin{document}

\flushbottom
\maketitle

\thispagestyle{empty}

\section*{Introduction}

Two dimensional semiconductors are an important class of functional nanomaterials with promising electronic and mechanical properties for optoelectronic and thermoelectric applications. Monolayer transition metal dichalcogenides of composition \ce{AB_2} (A = Mo/W and B = S/Se/Te) have recently attracted a lot of attention for optoelectronic properties arising from their favorable electronic band gaps in the range of 1.0 - 2.0 eV, high charge-carrier mobilities and large on/off ratios \cite{kumar2012electronic, lin2013modulating, sarkar2014mos2, wang2012electronics}. Thermal engineering of these monolayered materials remains a challenge for the design of devices based on two-dimensional materials. For instance, materials for thermal barrier coatings and thermoelectric energy generation require tight control over phonon transport over a wide range of frequencies to achieve minimal thermal conductivities \cite{sahoo2013temperature}, whereas materials for optoelectronic devices, where thermal dissipation is key, have opposing design requirements.\cite{peng2016thermal} Extensive efforts have been made to develop monolayered materials for thermoelectric applications, where a low lattice thermal conductivity is essential for achieving a high figure of merit \cite{RuanPRB2014, QuinnNature2001, LaForgeScience2002}. While several two-dimensional and layered materials have been characterized experimentally and computationally for their thermal transport properties \cite{LinNanoscale2019,TaishanNL2016,KanatzidisMRSB2015}, a systematic understanding of the role of point and extended defects and interfaces on controlling thermal conductivity in these systems is lacking.

However, several previous experimental and theoretical investigations have attempted to modulate lattice thermal transport in these material systems by a combination of alloying, interfacial and microstructural engineering and phase patterning. Alloying modifies thermal transport in materials by affecting one or more of the following material parameters -- crystal structure, atomic mass \cite{SinnottJACerS2009}, inter-atomic bonding and anharmonicity \cite{slack1973nonmetallic, lindsay2013first} and is effective in scattering high-frequency phonons \cite{sahoo2013temperature}. Formation of interfaces and superlattice structures in nanomaterials are very promising for controlling phonon scattering, particularly for low frequency phonons over 1-2 THz \cite{mak2010atomically, splendiani2010emerging, cao2012valley, wang2012integrated, ChengIJMT2015}. Scale-invariant fractal patterning, which results in features of multiple sizes, are widely pursued to affect phonons over a wide range of frequencies and mean free paths \cite{HanFractal2020}. These panoscopic techniques for hierarchical-design have been applied to identify electron-crystal and phonon-glass materials with excellent thermoelectric properties \cite{guo2019conformal}.

In this study, we use non-equilibrium molecular dynamics simulations (Section \ref{sec:NEMD}) to compute lattice thermal conductivity of monolayer \ce{(Mo|W)Se_2} systems, including \ce{Mo_{1-x}W_xSe_2} alloys (Section \ref{sec:alloy}) and fractal heterostructures (Section \ref{sec:fractal}) and periodic superlattices (Section \ref{sec:superlattice}) constructed out of two transition metal dichalcogenides, \ce{MoSe_2} and \ce{WSe_2}, suitable for ultra-thin electronic applications. This distribution of point defects, hetero-phase interfaces and a range of feature sizes allows us to explore the influence of each of these features on phonon scattering and identify guidelines for deisgn of two-dimensional material structures with tunable thermal transport properties.

\section*{Results}

\subsection*{Non-Equilibrium Molecular Dynamics Simulations for Computing Thermal Conductivity of \ce{(Mo|W)Se_2} Layers}
\label{sec:NEMD}
\begin{figure}
  \centering
  \includegraphics[width=\textwidth]{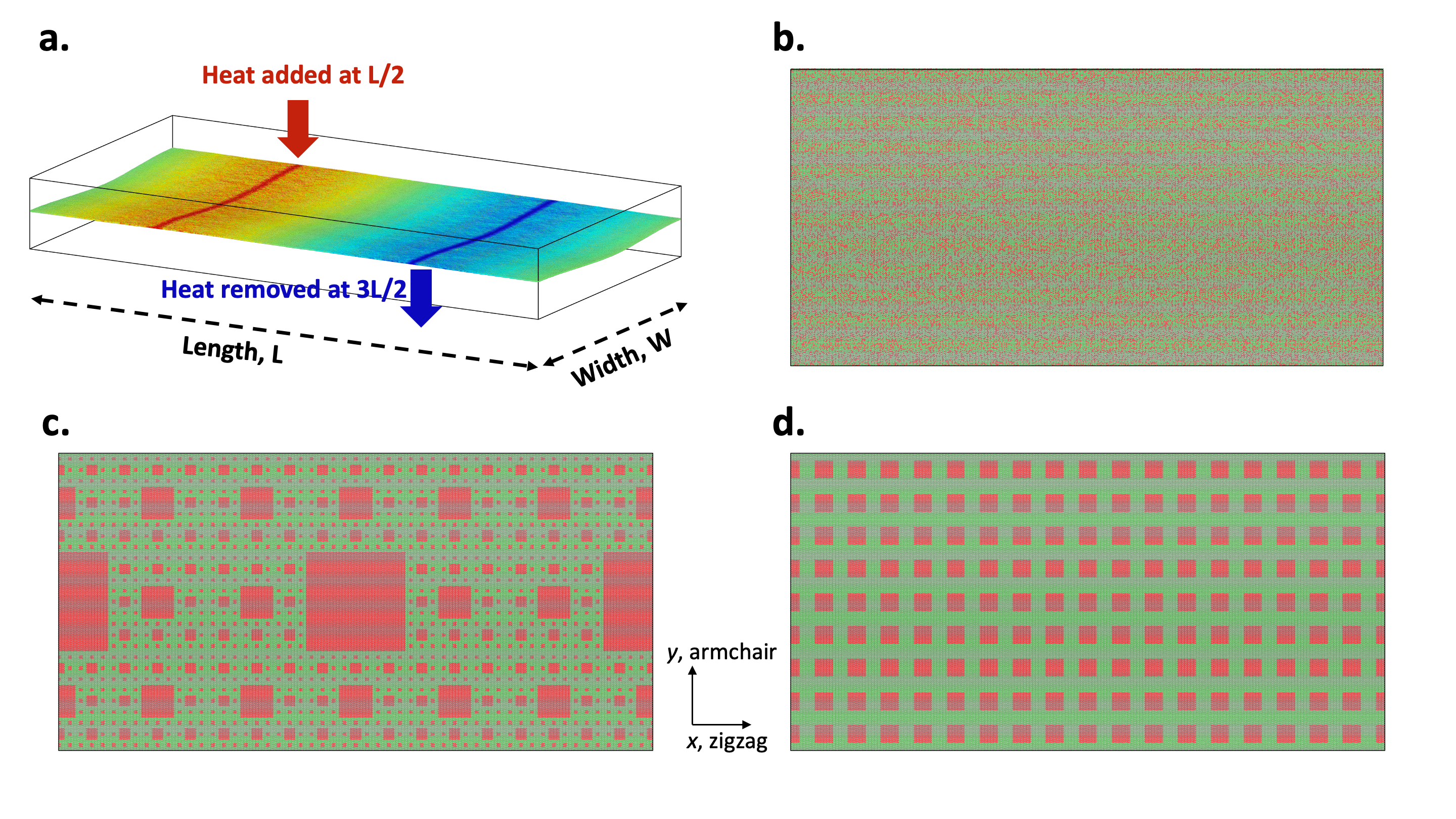}
  \caption{(a) Schematic of the thermal conductivity simulation. Heat is added at L/2 and removed at 3L/2 establishing a thermal gradient between the hot and cold regions. (b-d) Different (Mo$|$W)Se$_2$ systems with different phonon scattering features. Figure (b) shows a random distribution of \ce{WSe_2} (red) in a MoSe$_2$ (green) system, corresponding to a 40\% distribution of WSe$_2$ in MoSe$_2$. Figure c) shows a level 4 fractal heterostructure and (d) shows a periodic \ce{MoSe_2}$|$\ce{WSe_2} superlattice.}
  \label{fig:1}
\end{figure}

Lattice Thermal conductivity ($\kappa_{lat}$) of suspended monolayer crystals is computed using the so-called `direct' method of non-equilibrium molecular dynamics simulations (Figure \ref{fig:1}a). This non-perturbative approach for the calculation of $\kappa_{lat}$ for a heterogeneous system, is consistent with values extracted from classical equilibrium MD (EMD) simulations using Green-Kubo techniques \cite{SchellingNEMD2002}, but does not suffer from deficiencies in the commonly adopted relaxation time approximation solutions to the Boltzmann Transport Equation, which are known to severly underpredict the thermal conductivity of several 2D materials including transition metal dichalcogenides \cite{CepellottiPhononHydro2015,LindsayHBN2011}. To compute the $\kappa_{lat}$ for thermal transport along the $x$ direction in a \ce{(Mo|W)Se_2} monolayer of dimensions $2L \times L$, a predefined flux of thermal energy, $\dot{Q}$, is added to the atoms in a 100 \AA-strip at $x=\frac{L}{2}$ (`Hot' end) and an identical heat flux is removed from the system at $x=\frac{3L}{2}$ (`Cold' end). Periodic boundary conditions along the $x$- and $y$-directions, ensure an equal magnitude of thermal flux in the $x$ and $-x$ directions from the `Hot' to the `Cold' ends. The thermal conductivity of the system can then be obtained directly from the steady-state temperature gradient using the Fourier law of heat conduction (Equation \ref{eq:kappacalculate}).

\begin{equation}
  \kappa_{\mathrm{lattice}} = - \frac{1}{2\nabla T}\left[ \frac{\dot{Q}}{L\times t} \right]
  \label{eq:kappacalculate}
\end{equation}

where $\kappa_{\mathrm{lattice}}$ is the thermal conductivity of the monolayer, $\nabla T$ is the temperature gradient established between the heat source and heat sink due to the imposed heat flux, $\dot{Q}$. $L$ and $t$ are the effective width and thickness of the suspended monolayer. Thermal conductivity is calculated for four classes of \ce{(Mo|W)Se_2} systems containing different barriers to phonon propagation, namely, pure \ce{MoSe_2} and \ce{WSe_2} crystals with no point defects or interfaces, \ce{Mo_{1-x}W_xSe_2} substitutional alloys (Figure \ref{fig:1}b), self-similar fractal \ce{MoSe_2}/\ce{WSe_2} heterostructures (Figure \ref{fig:1}c), and periodic \ce{MoSe_2}/\ce{WSe_2} superlattices with a characteristic length scale, $l$ (Figure \ref{fig:1}d). The random \ce{Mo_{1-x}W_xSe_2} alloy is constructed by replacing $x$ fraction of cation sites chosen at random in the \ce{MoSe_2} lattice with \ce{W} atoms. Such a random alloy configuration is consistent with real TMDC alloys synthesized by scalable techniques like chemical vapor deposition (CVD) \cite{KochatAdvMater2017, ApteACSNano2018}. Periodic superlattices are constructed as a lattice of square \ce{WSe_2} patches of size $l$ in the \ce{MoSe_2} matrix separated by hetero-phase interfaces along the zigzag and armchair directions. Self-similar fractal structures are constructed by substitutionally alloying \ce{W} atoms in the cation sub-lattice of the \ce{MoSe_2} crystal in the form of a Sierpinski carpet. Results from these deterministic fractals are expected to hold even for random fractal structures of the same fractal dimension such as amorphous two-dimensional alloys \cite{spagnol2007thermal}. Both periodic superlattices and fractal heterostructures are constructed with atomically-sharp interfaces with no atomic mixing that can scatter short-wavelength phonons \cite{LAScience2012, LASA2018}. Such epitaxial interfaces between isoelectronic materials is preferable for optoelectronic applications, since diffuse interfaces, grain boundaries, inclusions and pores can also detrimentally affect electrical transport \cite{sahoo2013temperature}. Figure \ref{fig:1}c represents a representative fractal structures containing four levels of self-similarity. The choice of self-similarity level also dictates the overall stoichiometry of the fractal structure. All fractal structures are constructed such that the size of the smallest feature is larger than approximately 4 nm, reflecting the limits of current patterning technologies \cite{ChenPatterning2017}.

The average lattice strain in either the alloys or the heterostructures is less than -0.075\%, reflecting the near-identical in-plane lattice constants of \ce{MoSe_2} and \ce{WSe_2} ($a(\ce{MoSe_2}) = 3.289$\AA\ and $a(\ce{WSe_2}) = 3.286$\AA) \cite{BronsemaMoSe21986, SchutteWSe21987}. Therefore, point defects and interfacial scattering results mainly from changes in the bonding interactions and atomic masses and the potential effect of long-range disorder and strain on the measured thermal transport is negligible. Details about the molecular dynamics simulations, including development of suitable empirical forcefields and workflow are given in Section I and II of the Supporting Information.

\subsection*{Thermal transport in \ce{Mo_{1-x}W_xSe_2} alloys}
\label{sec:alloy}
\begin{figure}
  \centering
  \includegraphics[width=0.75\textwidth]{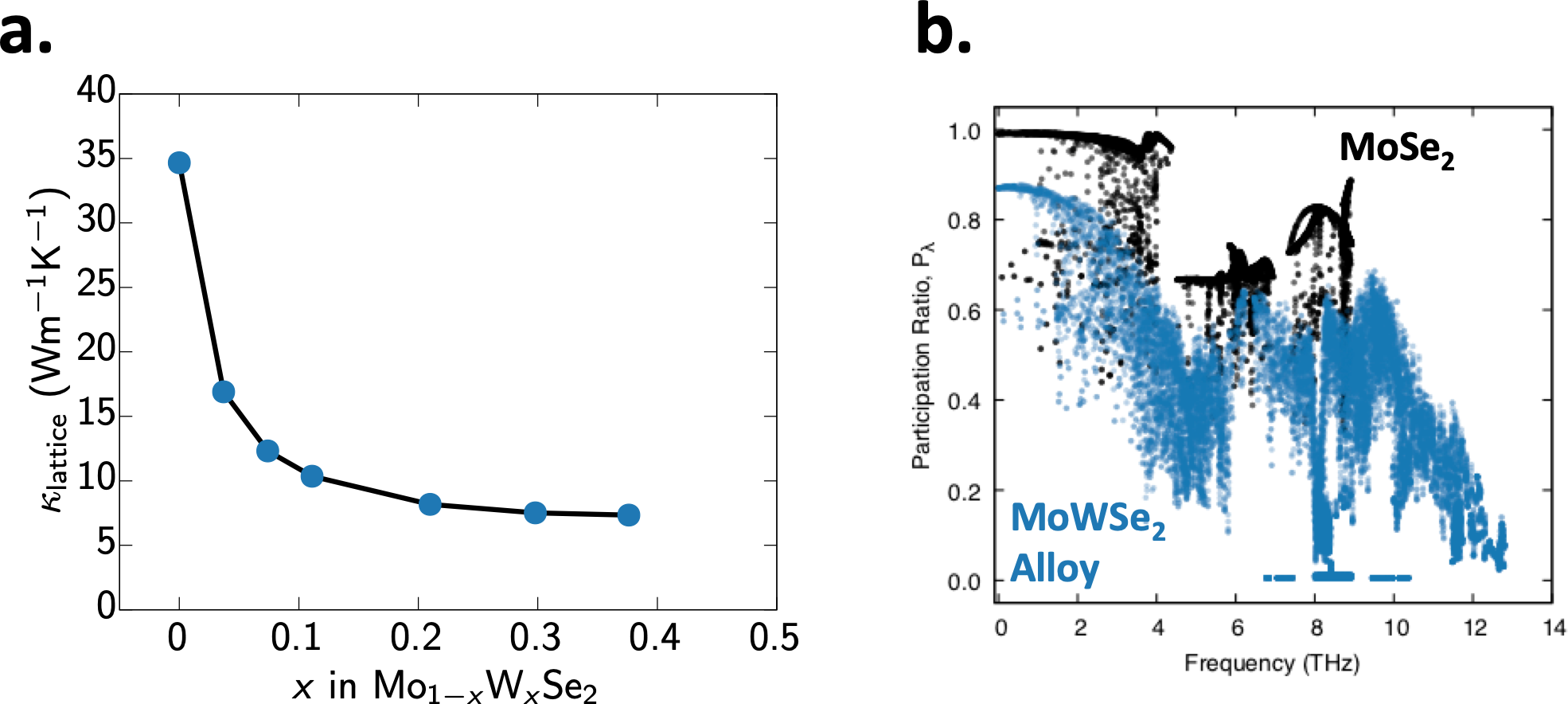}
  \caption{a) shows the variation of thermal conductivity with respect to the percentage of tungsten present in the system. b) The participation ratio of phonons in the pure and doped \ce{MoSe_2} crystals.}
  \label{fig:2}
\end{figure}

Substitutional doping of \ce{MoSe_2} by \ce{W} atoms has a significant effect on the lattice thermal conductivity. Figure \ref{fig:2}a shows the computed lattice thermal conductivity of the monolayer \ce{Mo_{1-x}W_xSe_2} alloy as a function of substitutional doping. Even moderate doping ($x < 5\%$) leads to greater than $70\%$ reduction in lattice thermal conductivity relative to undoped crystals. Similar results were observed in various materials\cite{abeles1963lattice, tian2012phonon, garg2011role, daly2002optical, LiAPL2009, XieRSCA2013}. Classical molecular dynamics simulations exclude electronic structure effects such as charge-transfer and charge carrier-phonon interactions, therefore the large reduction in $\kappa_{\mathrm{lattice}}$ is attributable primarily to increased rate of point defect scattering that originates from both the mass difference and inter-atomic coupling force differences resulting in greater phonon localization and reduced mean-free paths \cite{zhou2005influence, fleurial1997skutterudites, jung2017unusually}. However, there is no noticeable change in other phonon characteristics such as phonon frequencies, group velocities and phonon density of states at low frequencies. 

To quantify the phonon localization effect, we computed the phonon participation ratio $P_{\lambda}$ for the unalloyed and defect-free \ce{MoSe_2} single crystal and the 3.7\% W-doped \ce{MoSe_2} alloy (Figure \ref{fig:2}b). The phonon participation ratio, $P_{\lambda}$, measures the spatial localization of a phonon mode, $\lambda$ and it is defined as \cite{Keblinski06, ChenNanoscale2019}

\begin{equation}
P_{\lambda} = \frac{1}{N\sum_i \left(\sum_\alpha \varepsilon^*_{i\alpha,\lambda}\varepsilon_{i\alpha,\lambda}\right)^2}
\end{equation}

where $N$ is the total number of atoms and $\varepsilon_{i\alpha,\lambda}$ is the $\alpha^{th}$ cartesian component of the eigen-mode $\lambda$ for the $i$th atom. $P_{\lambda}$ is a dimensionless quantity ranging from $1/N$ to 1, with $\approx 1$ denoting the propagating mode and $\approx 0$ denoting the localized mode.

We observe that the degree of localization is enhanced for all phonons of finite frequency in doped \ce{MoSe_2} crystal, as shown by the lower values of $P_{\lambda}$ in doped-\ce{MoSe_2} as compared to that in dopant-free \ce{MoSe_2} single crystal samples. This behavior is consistent with Anderson's theory of localization of waves in disordered two-dimensional media driven by interference between multiple wave scattering \cite{AndersonPR58} as well as experimental observations in other two-dimensional materials \cite{RuanNL2014}. It can also be seen that substitutional point defects lead to a large suppression in thermal transport by high-frequency, low mean-free-path phonons, while long wavelength acoustic phonons undergo less scattering resulting in a finite and moderately large thermal conductivity even at high doping level. Further, it is noticeable that thermal conductivity of the alloy remains constant and relatively insensitive to \ce{W} content beyond approximately $20\%$ alloying. This low and composition-independent thermal conductivity implies that substitutional alloys are not suitable for thermal design applications.

\subsection*{Fractal \ce{MoSe_2}$|$\ce{WSe_2} heterostructures}
\label{sec:fractal}
\begin{figure}
  \centering
  \includegraphics[width=\textwidth]{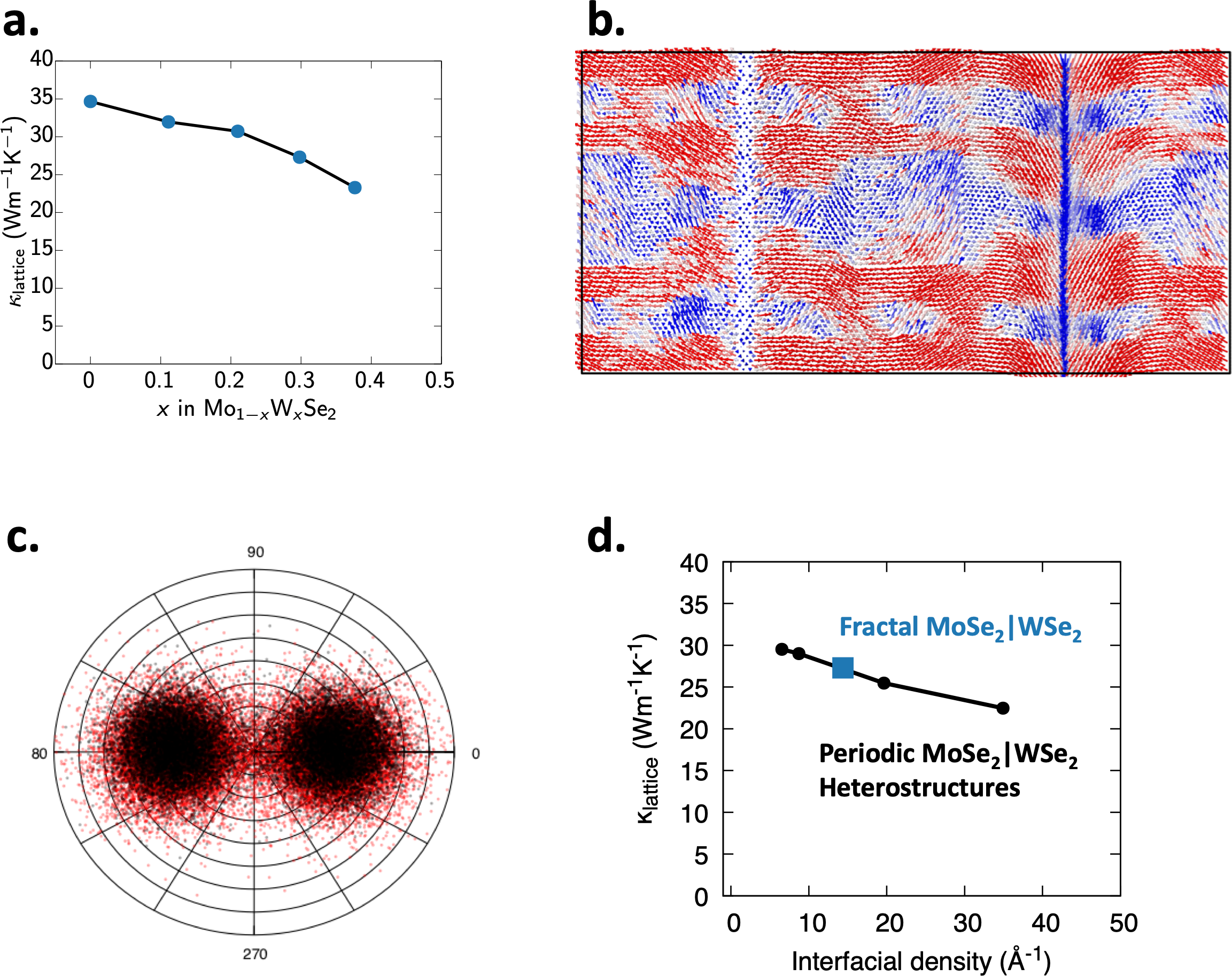}
  \caption{a) Graph showing variation of thermal conductivity in the self-similar \ce{MoSe_2}/\ce{WSe_2} heterostructure as a function total W content. b) Shows the per-atom heat flux vectors through the second-order fractal structure, with the arrows colored by the magnitude of the heat flux in the x-direction. It is apparent that the majority of the heat flux moves through the \ce{MoSe_2} lattice and the \ce{MoSe_2}/\ce{WSe_2} interface acts as the source of phonon scattering.(c) Angular distribution of local heat flux vectors in the pure (black) and heterostructured (red) \ce{MoSe_2} crystals. (d) Plot of thermal conductivity of periodic \ce{MoSe_2}/\ce{WSe_2} heterostructure as a function of thier interfacial density is consistent with the thermal conductivity of the fractal \ce{MoSe_2}$|$\ce{WSe_2} heterostructure, showing that incoherent phonons are the dominant thermal energy carriers in these materials.}
  \label{fig:3}
\end{figure}

There exist several empirical models to describe transport processes (electrical, thermal and mass) in porous, self-similar and fractal media \cite{ma2003self, ma2004fractal, miller1969bounds}. However, they provide a description of macroscopic properties of the system only in terms of the bulk properties of the individual phases, excluding any interfacial effects. The most common model for transport through irregular, porous and self-similar media is Archie's law \cite{archie1942electrical}. This empirical relation, given by $ \dot{Q} \propto \phi^m/a$ relates flux (thermal or mass) through the medium, $\dot{Q}$  to the phase fraction, $\phi$ and via the empirical exponent $m$ which takes a value between 1.3 - 2.5 and tortuosity of the thermal path, $a$\cite{thovert1990thermal}. An alternative model by Miller suggests that \cite{miller1969bounds}, $ \sigma_{max} = 1 - \left(\frac{1}{1-2G}\right)c $, where $c$ is the concentration of the \ce{WSe_2} phase (assumed to be of zero conductivity) and $\sigma_{max}$ is the thermal conductivity of the pure \ce{MoSe_2} phase and $G$ is some geometric parameter equal to 0.27 for square parches. Extending this thought, we can show that in a fractal of order $n$, the effective matrix around the largest central particle is a fractal of order $n-1$. Therefore, we can write $\sigma_{n} = \sigma_{n-1}* \left(1 - \frac{1}{1-2G}\right) $. This assumption is also common in more complex models for thermal transport in regular fractal systems. However, none of these models can accurately capture the gradual, near-linear variation of $\kappa_{lattice}$ with \ce{WSe_2} phase fraction, shown in Figure \ref{fig:3}a, because they do not consider the role of the \ce{MoSe_2}$|$\ce{WSe_2} interfacial scattering of phonons, which is the dominant scattering mechanism in these systems and the thermal boundary resistance of the \ce{MoSe_2}$|$\ce{WSe_2} interface, as described by the acoustic mismatch model \cite{PhilpottJAP2003}. Further, thermal transport in the resulting \ce{MoSe_2} and \ce{WSe_2} nano domains will also demonstrate significant size effects within the Casimir regime (i.e. smallest feature size $<$ phonon mean free path). Therefore Archie's law and other previously determined models cannot be applied, contrary to the results of Ref. \cite{thovert1990thermal}.

In these self-similar structures, the reduction in thermal conductivity is caused by phonon scattering at \ce{MoSe_2}/\ce{WSe_2} heterointerfaces. To understand this scattering process, we compute the time-averaged heat flux on each atom in NEMD simulations using the expression
\begin{equation}
  q = e \cdot v_i - S_{ij}\cdot v_j
\end{equation}
where $e$, $v_i$, and $S_{ij}$ are the energy, velocity vector, and local stress tensor at each atom \cite{XuAPL2011, SiMolecules2018}. Figure \ref{fig:3}b shows the computed per-atom flux through the fractal-patterned \ce{MoSe_2}/\ce{WSe_2} heterostructure. It is noticeable that the \ce{MoSe_2}/\ce{WSe_2} interfaces are the primary source of phonon scattering and that the majority of the thermal flux flows through regions of the fractal structure that contain no \ce{MoSe_2}/\ce{WSe_2} interfaces in the $x$-direction. The figure also shows that the majority of the thermal boundary resistance is concentrated at the interfaces closest to the hot or the cold end, consistent with observations from the Si-Ge system \cite{RanGeSi2018}.

\subsection*{Thermal transport in periodic superlattices}
\label{sec:superlattice}
In order to understand if the inherent lack of periodicity in the fractal structure affects phonon propagation, we also compute the thermal conductivity of periodic \ce{MoSe_2}$|$\ce{WSe_2} superlattices with square patches of \ce{WSe_2} patches embedded in a \ce{MoSe_2} matrix (Figure \ref{fig:1}d). Specifically, we choose heterostructures of composition 29\% \ce{WSe_2}, equal to that in a level 3 fractal heterostructure, for our simulations. At this constant composition, we can vary the periodicity of \ce{WSe_2} patches to construct periodic heterostructures of different interfacial densities.

Figure \ref{fig:3}d shows the near-linear decrease in the computed thermal conductivity of the three periodic heterostructures as a function of interfacial density, as seen in other semiconducting systems like Si-Ge \cite{MajumdarJHT2002, ChengIJMT2015}. It can be observed that the computed $\kappa_{\mathrm{lattice}}$ for the third-level fractal falls in line with the trend predicted by the periodic heterostructures. This linear and inverse dependence of thermal resistance with interfacial density (and not by their relative orientations and arrangement) in indicates that thermal transport in \ce{MoSe_2}/\ce{WSe_2} heterostructures is dominated by conduction of incoherent phonons. The presence of interfaces and anharmonicity of the interatomic interactions lead to decoherence of phonons and their resulting particle-like behavior \cite{ZurbuchenNMat2014}. Coherent phonons, which can traverse periodic heterostructure, but not non-periodic fractal ones \cite{ChenNanoscale2019}, contribute negligibly to the calculated thermal conductivity.

\subsection*{Design of heterostructures for tuning lattice thermal transport}

\begin{figure}
  \centering
  \includegraphics[width=\textwidth]{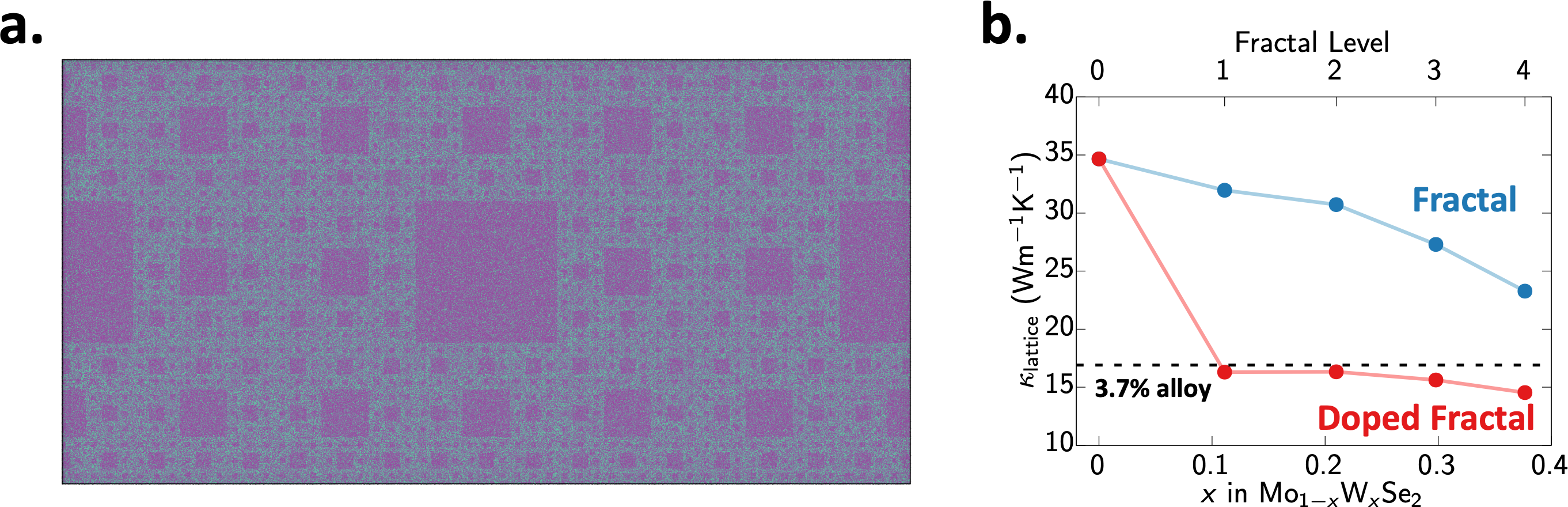}
  \caption{a) Configuration of alloyed fractal structure with 3.7\% \ce{W} alloying. b) The combination of two distinct phonon scattering mechanisms (i.e. interfacial formation and point defect scattering) results in a lower thermal conductivity for the alloyed fractal system (red) than cab be achieved in either pure 3.7\% alloying (black line) or the undoped fractals (blue)}
  \label{fig:4}
\end{figure}

This understanding of phonon scattering by point defects (like vacancies and dopant atoms) and heterostructure interfaces provides useful design guidelines for the construction of low thermal conductivity structure. Figure \ref{fig:4}a shows one such heterostructure which attempts to maximize both the interfacial density as well as the concentration of dopant atoms in the \ce{WSe_2} patches and the \ce{MoSe_2} matrix. This `doped fractal' structure was observed to have a thermal conductivity of only 15 W/mK, which is lower than that of either the 3\%-doped \ce{Mo_{1-x}W_xSe_2} alloy or the third-level fractal \ce{MoSe_2}$|$\ce{WSe_2} heterostructure used to construct the `doped' fractal structure(Figure \ref{fig:4}b). This behavior can be explained using Matthiessen's rule of independent scattering events, where the overall scattering rate is a sum of individual scattering rates \cite{XieRSCA2013}. These simulations show that careful control over doping and heterostructure construction can be used to controllably modify thermal conductivity of \ce{(Mo|W)Se_2} monolayer single crystals.

\section*{Discussion}

We have performed non-equilibrium molecular dynamics simulations using a specifically parameterized force-field to compare the thermal conductivity of suspended \ce{Mo_{1-x}W_xSe_2} alloys with periodic and fractal-patterned \ce{MoSe_2}$|$\ce{WSe_2} heterostructures to identify the dependence of lattice thermal conductivity on dopant concentrations and interfacial densities. We show that even low dopant concentrations ($<$ 5\% doping) can strongly localize high-frequency phonons in the \ce{(Mo|W)Se_2} crystal leading to a large ($>$ 70\%) reduction in the lattice thermal conductivity. Further, this low value of $\kappa_{\mathrm{lattice}}$ is largely insensitive to dopant concentration and therefore alloying alone is not a viable strategy for controlling thermal conductivity. On the other hand, thermal transport in both periodic and fractal patterned heterostructures is dominated by incoherent phonon conduction and varies gradually and monotonically with the density of \ce{MoSe_2}$|$\ce{WSe_2} interfaces. Thermal conductivity can be controllably tuned by constructign doped fractal heterostructures where both scattering mechanisms operate.

\bibliography{biblist.bib}

\begin{thebibliography}{10}
\urlstyle{rm}
\expandafter\ifx\csname url\endcsname\relax
  \def\url#1{\texttt{#1}}\fi
\expandafter\ifx\csname urlprefix\endcsname\relax\def\urlprefix{URL }\fi
\expandafter\ifx\csname doiprefix\endcsname\relax\def\doiprefix{DOI: }\fi
\providecommand{\bibinfo}[2]{#2}
\providecommand{\eprint}[2][]{\url{#2}}

\bibitem{kumar2012electronic}
\bibinfo{author}{Kumar, A.} \& \bibinfo{author}{Ahluwalia, P.}
\newblock \bibinfo{journal}{\bibinfo{title}{Electronic structure of transition
  metal dichalcogenides monolayers 1h-mx 2 (m= mo, w; x= s, se, te) from
  ab-initio theory: new direct band gap semiconductors}}.
\newblock {\emph{\JournalTitle{The European Physical Journal B}}}
  \textbf{\bibinfo{volume}{85}}, \bibinfo{pages}{186} (\bibinfo{year}{2012}).

\bibitem{lin2013modulating}
\bibinfo{author}{Lin, J.} \emph{et~al.}
\newblock \bibinfo{journal}{\bibinfo{title}{Modulating electronic transport
  properties of mos2 field effect transistor by surface overlayers}}.
\newblock {\emph{\JournalTitle{Applied Physics Letters}}}
  \textbf{\bibinfo{volume}{103}}, \bibinfo{pages}{063109}
  (\bibinfo{year}{2013}).

\bibitem{sarkar2014mos2}
\bibinfo{author}{Sarkar, D.} \emph{et~al.}
\newblock \bibinfo{journal}{\bibinfo{title}{Mos2 field-effect transistor for
  next-generation label-free biosensors}}.
\newblock {\emph{\JournalTitle{ACS nano}}} \textbf{\bibinfo{volume}{8}},
  \bibinfo{pages}{3992--4003} (\bibinfo{year}{2014}).

\bibitem{wang2012electronics}
\bibinfo{author}{Wang, Q.~H.}, \bibinfo{author}{Kalantar-Zadeh, K.},
  \bibinfo{author}{Kis, A.}, \bibinfo{author}{Coleman, J.~N.} \&
  \bibinfo{author}{Strano, M.~S.}
\newblock \bibinfo{journal}{\bibinfo{title}{Electronics and optoelectronics of
  two-dimensional transition metal dichalcogenides}}.
\newblock {\emph{\JournalTitle{Nature nanotechnology}}}
  \textbf{\bibinfo{volume}{7}}, \bibinfo{pages}{699--712}
  (\bibinfo{year}{2012}).

\bibitem{sahoo2013temperature}
\bibinfo{author}{Sahoo, S.}, \bibinfo{author}{Gaur, A.~P.},
  \bibinfo{author}{Ahmadi, M.}, \bibinfo{author}{Guinel, M. J.-F.} \&
  \bibinfo{author}{Katiyar, R.~S.}
\newblock \bibinfo{journal}{\bibinfo{title}{Temperature-dependent raman studies
  and thermal conductivity of few-layer mos2}}.
\newblock {\emph{\JournalTitle{The Journal of Physical Chemistry C}}}
  \textbf{\bibinfo{volume}{117}}, \bibinfo{pages}{9042--9047}
  (\bibinfo{year}{2013}).

\bibitem{peng2016thermal}
\bibinfo{author}{Peng, B.} \emph{et~al.}
\newblock \bibinfo{journal}{\bibinfo{title}{Thermal conductivity of monolayer
  mos 2, mose 2, and ws 2: interplay of mass effect, interatomic bonding and
  anharmonicity}}.
\newblock {\emph{\JournalTitle{RSC Advances}}} \textbf{\bibinfo{volume}{6}},
  \bibinfo{pages}{5767--5773} (\bibinfo{year}{2016}).

\bibitem{RuanPRB2014}
\bibinfo{author}{Wang, Y.}, \bibinfo{author}{Huang, H.~X.} \&
  \bibinfo{author}{Ruan, X.~L.}
\newblock \bibinfo{journal}{\bibinfo{title}{Decomposition of coherent and
  incoherent phonon conduction in superlattices and random multilayers}}.
\newblock {\emph{\JournalTitle{Physical Review B}}}
  \textbf{\bibinfo{volume}{90}}, \doiprefix\url{10.1103/PhysRevB.90.165406}
  (\bibinfo{year}{2014}).

\bibitem{QuinnNature2001}
\bibinfo{author}{Venkatasubramanian, R.}, \bibinfo{author}{Siivola, E.},
  \bibinfo{author}{Colpitts, T.} \& \bibinfo{author}{O'Quinn, B.}
\newblock \bibinfo{journal}{\bibinfo{title}{Thin-film thermoelectric devices
  with high room-temperature figures of merit}}.
\newblock {\emph{\JournalTitle{Nature}}} \textbf{\bibinfo{volume}{413}},
  \bibinfo{pages}{597--602}, \doiprefix\url{Doi 10.1038/35098012}
  (\bibinfo{year}{2001}).

\bibitem{LaForgeScience2002}
\bibinfo{author}{Harman, T.~C.}, \bibinfo{author}{Taylor, P.~J.},
  \bibinfo{author}{Walsh, M.~P.} \& \bibinfo{author}{LaForge, B.~E.}
\newblock \bibinfo{journal}{\bibinfo{title}{Quantum dot superlattice
  thermoelectric materials and devices}}.
\newblock {\emph{\JournalTitle{Science}}} \textbf{\bibinfo{volume}{297}},
  \bibinfo{pages}{2229--2232}, \doiprefix\url{DOI 10.1126/science.1072886}
  (\bibinfo{year}{2002}).

\bibitem{LinNanoscale2019}
\bibinfo{author}{Han, D.}, \bibinfo{author}{Ding, W.}, \bibinfo{author}{Wang,
  X.} \& \bibinfo{author}{Cheng, L.}
\newblock \bibinfo{journal}{\bibinfo{title}{Tunable thermal transport in a ws2
  monolayer with isotopic doping and fractal structure}}.
\newblock {\emph{\JournalTitle{Nanoscale}}} \textbf{\bibinfo{volume}{11}},
  \bibinfo{pages}{19763--19771}, \doiprefix\url{10.1039/C9NR02835H}
  (\bibinfo{year}{2019}).

\bibitem{TaishanNL2016}
\bibinfo{author}{Zhu, T.} \& \bibinfo{author}{Ertekin, E.}
\newblock \bibinfo{journal}{\bibinfo{title}{Phonons, localization, and thermal
  conductivity of diamond nanothreads and amorphous graphene}}.
\newblock {\emph{\JournalTitle{Nano Letters}}} \textbf{\bibinfo{volume}{16}},
  \bibinfo{pages}{4763--4772}, \doiprefix\url{10.1021/acs.nanolett.6b00557}
  (\bibinfo{year}{2016}).
\newblock \bibinfo{note}{PMID: 27388115}.

\bibitem{KanatzidisMRSB2015}
\bibinfo{author}{Kanatzidis, M.~G.}
\newblock \bibinfo{journal}{\bibinfo{title}{Advances in thermoelectrics: From
  single phases to hierarchical nanostructures and back}}.
\newblock {\emph{\JournalTitle{MRS Bulletin}}} \textbf{\bibinfo{volume}{40}},
  \bibinfo{pages}{687–695}, \doiprefix\url{10.1557/mrs.2015.173}
  (\bibinfo{year}{2015}).

\bibitem{SinnottJACerS2009}
\bibinfo{author}{Watanabe, T.} \emph{et~al.}
\newblock \bibinfo{journal}{\bibinfo{title}{Thermal transport in
  off-stoichiometric uranium dioxide by atomic level simulation}}.
\newblock {\emph{\JournalTitle{Journal of the American Ceramic Society}}}
  \textbf{\bibinfo{volume}{92}}, \bibinfo{pages}{850--856},
  \doiprefix\url{10.1111/j.1551-2916.2009.02966.x} (\bibinfo{year}{2009}).
\newblock
  \eprint{https://ceramics.onlinelibrary.wiley.com/doi/pdf/10.1111/j.1551-2916.2009.02966.x}.

\bibitem{slack1973nonmetallic}
\bibinfo{author}{Slack, G.~A.}
\newblock \bibinfo{journal}{\bibinfo{title}{Nonmetallic crystals with high
  thermal conductivity}}.
\newblock {\emph{\JournalTitle{Journal of Physics and Chemistry of Solids}}}
  \textbf{\bibinfo{volume}{34}}, \bibinfo{pages}{321--335}
  (\bibinfo{year}{1973}).

\bibitem{lindsay2013first}
\bibinfo{author}{Lindsay, L.}, \bibinfo{author}{Broido, D.} \&
  \bibinfo{author}{Reinecke, T.}
\newblock \bibinfo{journal}{\bibinfo{title}{First-principles determination of
  ultrahigh thermal conductivity of boron arsenide: A competitor for diamond?}}
\newblock {\emph{\JournalTitle{Physical review letters}}}
  \textbf{\bibinfo{volume}{111}}, \bibinfo{pages}{025901}
  (\bibinfo{year}{2013}).

\bibitem{mak2010atomically}
\bibinfo{author}{Mak, K.~F.}, \bibinfo{author}{Lee, C.}, \bibinfo{author}{Hone,
  J.}, \bibinfo{author}{Shan, J.} \& \bibinfo{author}{Heinz, T.~F.}
\newblock \bibinfo{journal}{\bibinfo{title}{Atomically thin mos 2: a new
  direct-gap semiconductor}}.
\newblock {\emph{\JournalTitle{Physical review letters}}}
  \textbf{\bibinfo{volume}{105}}, \bibinfo{pages}{136805}
  (\bibinfo{year}{2010}).

\bibitem{splendiani2010emerging}
\bibinfo{author}{Splendiani, A.} \emph{et~al.}
\newblock \bibinfo{journal}{\bibinfo{title}{Emerging photoluminescence in
  monolayer mos2}}.
\newblock {\emph{\JournalTitle{Nano letters}}} \textbf{\bibinfo{volume}{10}},
  \bibinfo{pages}{1271--1275} (\bibinfo{year}{2010}).

\bibitem{cao2012valley}
\bibinfo{author}{Cao, T.} \emph{et~al.}
\newblock \bibinfo{journal}{\bibinfo{title}{Valley-selective circular dichroism
  of monolayer molybdenum disulphide}}.
\newblock {\emph{\JournalTitle{Nature communications}}}
  \textbf{\bibinfo{volume}{3}}, \bibinfo{pages}{887} (\bibinfo{year}{2012}).

\bibitem{wang2012integrated}
\bibinfo{author}{Wang, H.} \emph{et~al.}
\newblock \bibinfo{journal}{\bibinfo{title}{Integrated circuits based on
  bilayer mos2 transistors}}.
\newblock {\emph{\JournalTitle{Nano letters}}} \textbf{\bibinfo{volume}{12}},
  \bibinfo{pages}{4674--4680} (\bibinfo{year}{2012}).

\bibitem{ChengIJMT2015}
\bibinfo{author}{Chen, Y.~P.}, \bibinfo{author}{Deng, Z.~L.} \&
  \bibinfo{author}{Cheng, Q.~K.}
\newblock \bibinfo{journal}{\bibinfo{title}{Thermal conductivity of si/ge
  nanocomposites with fractal tree-shaped networks by considering the phonon
  interface scattering}}.
\newblock {\emph{\JournalTitle{International Journal of Heat and Mass
  Transfer}}} \textbf{\bibinfo{volume}{88}}, \bibinfo{pages}{572--578},
  \doiprefix\url{10.1016/j.ijheatmasstransfer.2015.04.093}
  (\bibinfo{year}{2015}).

\bibitem{HanFractal2020}
\bibinfo{author}{Han, D.}, \bibinfo{author}{Fan, H.}, \bibinfo{author}{Wang,
  X.} \& \bibinfo{author}{Cheng, L.}
\newblock \bibinfo{journal}{\bibinfo{title}{Atomistic simulations of phonon
  behaviors in isotopically doped graphene with sierpinski carpet fractal
  structure}}.
\newblock {\emph{\JournalTitle{Materials Research Express}}}
  \textbf{\bibinfo{volume}{7}}, \bibinfo{pages}{035020},
  \doiprefix\url{10.1088/2053-1591/ab7e4b} (\bibinfo{year}{2020}).

\bibitem{guo2019conformal}
\bibinfo{author}{Guo, J.}, \bibinfo{author}{Yang, F.}, \bibinfo{author}{Xia,
  M.}, \bibinfo{author}{Xu, X.} \& \bibinfo{author}{Li, B.}
\newblock \bibinfo{journal}{\bibinfo{title}{Conformal interface of monolayer
  molybdenum diselenide/disulfide and dielectric substrate with improved
  thermal dissipation}}.
\newblock {\emph{\JournalTitle{Journal of Physics D: Applied Physics}}}
  (\bibinfo{year}{2019}).

\bibitem{SchellingNEMD2002}
\bibinfo{author}{Schelling, P.~K.}, \bibinfo{author}{Phillpot, S.~R.} \&
  \bibinfo{author}{Keblinski, P.}
\newblock \bibinfo{journal}{\bibinfo{title}{Comparison of atomic-level
  simulation methods for computing thermal conductivity}}.
\newblock {\emph{\JournalTitle{Phys. Rev. B}}} \textbf{\bibinfo{volume}{65}},
  \bibinfo{pages}{144306}, \doiprefix\url{10.1103/PhysRevB.65.144306}
  (\bibinfo{year}{2002}).

\bibitem{CepellottiPhononHydro2015}
\bibinfo{author}{Cepellotti, A.} \emph{et~al.}
\newblock \bibinfo{journal}{\bibinfo{title}{Phonon hydrodynamics in
  two-dimensional materials.}}
\newblock {\emph{\JournalTitle{Nature Communications}}}
  \textbf{\bibinfo{volume}{6}}, \bibinfo{pages}{6400} (\bibinfo{year}{2015}).

\bibitem{LindsayHBN2011}
\bibinfo{author}{Lindsay, L.} \& \bibinfo{author}{Broido, D.~A.}
\newblock \bibinfo{journal}{\bibinfo{title}{Enhanced thermal conductivity and
  isotope effect in single-layer hexagonal boron nitride}}.
\newblock {\emph{\JournalTitle{Phys. Rev. B}}} \textbf{\bibinfo{volume}{84}},
  \bibinfo{pages}{155421}, \doiprefix\url{10.1103/PhysRevB.84.155421}
  (\bibinfo{year}{2011}).

\bibitem{KochatAdvMater2017}
\bibinfo{author}{Kochat, V.} \emph{et~al.}
\newblock \bibinfo{journal}{\bibinfo{title}{Re doping in 2d transition metal
  dichalcogenides as a new route to tailor structural phases and induced
  magnetism}}.
\newblock {\emph{\JournalTitle{Advanced Materials}}}
  \textbf{\bibinfo{volume}{29}}, \bibinfo{pages}{1703754},
  \doiprefix\url{10.1002/adma.201703754} (\bibinfo{year}{2017}).

\bibitem{ApteACSNano2018}
\bibinfo{author}{Apte, A.} \emph{et~al.}
\newblock \bibinfo{journal}{\bibinfo{title}{Structural phase transformation in
  strained monolayer mowse2 alloy}}.
\newblock {\emph{\JournalTitle{ACS Nano}}} \textbf{\bibinfo{volume}{12}},
  \bibinfo{pages}{3468--3476}, \doiprefix\url{10.1021/acsnano.8b00248}
  (\bibinfo{year}{2018}).

\bibitem{spagnol2007thermal}
\bibinfo{author}{Spagnol, S.}, \bibinfo{author}{Lartigue, B.},
  \bibinfo{author}{Trombe, A.} \& \bibinfo{author}{Gibiat, V.}
\newblock \bibinfo{journal}{\bibinfo{title}{Thermal modeling of two-dimensional
  periodic fractal patterns, an application to nanoporous media}}.
\newblock {\emph{\JournalTitle{EPL (Europhysics Letters)}}}
  \textbf{\bibinfo{volume}{78}}, \bibinfo{pages}{46005} (\bibinfo{year}{2007}).

\bibitem{LAScience2012}
\bibinfo{author}{Luckyanova, M.~N.} \emph{et~al.}
\newblock \bibinfo{journal}{\bibinfo{title}{Coherent phonon heat conduction in
  superlattices}}.
\newblock {\emph{\JournalTitle{Science}}} \textbf{\bibinfo{volume}{338}},
  \bibinfo{pages}{936--939}, \doiprefix\url{10.1126/science.1225549}
  (\bibinfo{year}{2012}).

\bibitem{LASA2018}
\bibinfo{author}{Luckyanova, M.~N.} \emph{et~al.}
\newblock \bibinfo{journal}{\bibinfo{title}{Phonon localization in heat
  conduction}}.
\newblock {\emph{\JournalTitle{Science Advances}}}
  \textbf{\bibinfo{volume}{4}}, \doiprefix\url{10.1126/sciadv.aat9460}
  (\bibinfo{year}{2018}).

\bibitem{ChenPatterning2017}
\bibinfo{author}{Chen, M.}, \bibinfo{author}{Rokni, H.}, \bibinfo{author}{Lu,
  W.} \& \bibinfo{author}{Liang, X.}
\newblock \bibinfo{journal}{\bibinfo{title}{Scaling behavior of nanoimprint and
  nanoprinting lithography for producing nanostructures of molybdenum
  disulfide}}.
\newblock {\emph{\JournalTitle{Microsystems \& Nanoengineering}}}
  \textbf{\bibinfo{volume}{3}}, \bibinfo{pages}{micronano201753},
  \doiprefix\url{10.1038/micronano.2017.53} (\bibinfo{year}{2017}).

\bibitem{BronsemaMoSe21986}
\bibinfo{author}{Bronsema, K.~D.}, \bibinfo{author}{De~Boer, J.~L.} \&
  \bibinfo{author}{Jellinek, F.}
\newblock \bibinfo{journal}{\bibinfo{title}{On the structure of molybdenum
  diselenide and disulfide}}.
\newblock {\emph{\JournalTitle{Zeitschrift fur anorganische und allgemeine
  Chemie}}} \textbf{\bibinfo{volume}{540}}, \bibinfo{pages}{15--17},
  \doiprefix\url{10.1002/zaac.19865400904} (\bibinfo{year}{1986}).
\newblock
  \eprint{https://onlinelibrary.wiley.com/doi/pdf/10.1002/zaac.19865400904}.

\bibitem{SchutteWSe21987}
\bibinfo{author}{Schutte, W.}, \bibinfo{author}{Boer, J.~D.} \&
  \bibinfo{author}{Jellinek, F.}
\newblock \bibinfo{journal}{\bibinfo{title}{Crystal structures of tungsten
  disulfide and diselenide}}.
\newblock {\emph{\JournalTitle{Journal of Solid State Chemistry}}}
  \textbf{\bibinfo{volume}{70}}, \bibinfo{pages}{207 -- 209},
  \doiprefix\url{10.1016/0022-4596(87)90057-0} (\bibinfo{year}{1987}).

\bibitem{abeles1963lattice}
\bibinfo{author}{Abeles, B.}
\newblock \bibinfo{journal}{\bibinfo{title}{Lattice thermal conductivity of
  disordered semiconductor alloys at high temperatures}}.
\newblock {\emph{\JournalTitle{Physical Review}}}
  \textbf{\bibinfo{volume}{131}}, \bibinfo{pages}{1906} (\bibinfo{year}{1963}).

\bibitem{tian2012phonon}
\bibinfo{author}{Tian, Z.} \emph{et~al.}
\newblock \bibinfo{journal}{\bibinfo{title}{Phonon conduction in pbse, pbte,
  and pbte 1- x se x from first-principles calculations}}.
\newblock {\emph{\JournalTitle{Physical Review B}}}
  \textbf{\bibinfo{volume}{85}}, \bibinfo{pages}{184303}
  (\bibinfo{year}{2012}).

\bibitem{garg2011role}
\bibinfo{author}{Garg, J.}, \bibinfo{author}{Bonini, N.},
  \bibinfo{author}{Kozinsky, B.} \& \bibinfo{author}{Marzari, N.}
\newblock \bibinfo{journal}{\bibinfo{title}{Role of disorder and anharmonicity
  in the thermal conductivity of silicon-germanium alloys: A first-principles
  study}}.
\newblock {\emph{\JournalTitle{Physical review letters}}}
  \textbf{\bibinfo{volume}{106}}, \bibinfo{pages}{045901}
  (\bibinfo{year}{2011}).

\bibitem{daly2002optical}
\bibinfo{author}{Daly, B.}, \bibinfo{author}{Maris, H.},
  \bibinfo{author}{Nurmikko, A.}, \bibinfo{author}{Kuball, M.} \&
  \bibinfo{author}{Han, J.}
\newblock \bibinfo{journal}{\bibinfo{title}{Optical pump-and-probe measurement
  of the thermal conductivity of nitride thin films}}.
\newblock {\emph{\JournalTitle{Journal of applied physics}}}
  \textbf{\bibinfo{volume}{92}}, \bibinfo{pages}{3820--3824}
  (\bibinfo{year}{2002}).

\bibitem{LiAPL2009}
\bibinfo{author}{Chen, J.}, \bibinfo{author}{Zhang, G.} \& \bibinfo{author}{Li,
  B.~W.}
\newblock \bibinfo{journal}{\bibinfo{title}{Tunable thermal conductivity of
  si1-xgex nanowires}}.
\newblock {\emph{\JournalTitle{Applied Physics Letters}}}
  \textbf{\bibinfo{volume}{95}}, \doiprefix\url{10.1063/1.3212737}
  (\bibinfo{year}{2009}).

\bibitem{XieRSCA2013}
\bibinfo{author}{Wang, Y.~C.}, \bibinfo{author}{Li, B.~H.} \&
  \bibinfo{author}{Xie, G.~F.}
\newblock \bibinfo{journal}{\bibinfo{title}{Significant reduction of thermal
  conductivity in silicon nanowires by shell doping}}.
\newblock {\emph{\JournalTitle{{RSC} Advances}}} \textbf{\bibinfo{volume}{3}},
  \bibinfo{pages}{26074--26079}, \doiprefix\url{10.1039/c3ra45113e}
  (\bibinfo{year}{2013}).

\bibitem{zhou2005influence}
\bibinfo{author}{Zhou, Z.}, \bibinfo{author}{Uher, C.},
  \bibinfo{author}{Jewell, A.} \& \bibinfo{author}{Caillat, T.}
\newblock \bibinfo{journal}{\bibinfo{title}{Influence of point-defect
  scattering on the lattice thermal conductivity of solid solution co (sb 1- x
  as x) 3}}.
\newblock {\emph{\JournalTitle{Physical Review B}}}
  \textbf{\bibinfo{volume}{71}}, \bibinfo{pages}{235209}
  (\bibinfo{year}{2005}).

\bibitem{fleurial1997skutterudites}
\bibinfo{author}{Fleurial, J.-P.}, \bibinfo{author}{Caillat, T.} \&
  \bibinfo{author}{Borshchevsky, A.}
\newblock \bibinfo{title}{Skutterudites: an update}.
\newblock In \emph{\bibinfo{booktitle}{Thermoelectrics, 1997. Proceedings
  ICT'97. XVI International Conference on}}, \bibinfo{pages}{1--11}
  (\bibinfo{organization}{IEEE}, \bibinfo{year}{1997}).

\bibitem{jung2017unusually}
\bibinfo{author}{Jung, G.~S.}, \bibinfo{author}{Yeo, J.},
  \bibinfo{author}{Tian, Z.}, \bibinfo{author}{Qin, Z.} \&
  \bibinfo{author}{Buehler, M.~J.}
\newblock \bibinfo{journal}{\bibinfo{title}{Unusually low and
  density-insensitive thermal conductivity of three-dimensional gyroid
  graphene}}.
\newblock {\emph{\JournalTitle{Nanoscale}}} \textbf{\bibinfo{volume}{9}},
  \bibinfo{pages}{13477--13484} (\bibinfo{year}{2017}).

\bibitem{Keblinski06}
\bibinfo{author}{Bodapati, A.}, \bibinfo{author}{Schelling, P.~K.},
  \bibinfo{author}{Phillpot, S.~R.} \& \bibinfo{author}{Keblinski, P.}
\newblock \bibinfo{journal}{\bibinfo{title}{Vibrations and thermal transport in
  nanocrystalline silicon}}.
\newblock {\emph{\JournalTitle{Physical Review B}}}
  \textbf{\bibinfo{volume}{74}}, \doiprefix\url{10.1103/PhysRevB.74.245207}
  (\bibinfo{year}{2006}).

\bibitem{ChenNanoscale2019}
\bibinfo{author}{Hu, S.~Q.} \emph{et~al.}
\newblock \bibinfo{journal}{\bibinfo{title}{Disorder limits the coherent phonon
  transport in two-dimensional phononic crystal structures}}.
\newblock {\emph{\JournalTitle{Nanoscale}}} \textbf{\bibinfo{volume}{11}},
  \bibinfo{pages}{11839--11846}, \doiprefix\url{10.1039/c9nr02548k}
  (\bibinfo{year}{2019}).

\bibitem{AndersonPR58}
\bibinfo{author}{Anderson, P.~W.}
\newblock \bibinfo{journal}{\bibinfo{title}{Absence of diffusion in certain
  random lattices}}.
\newblock {\emph{\JournalTitle{Physical Review}}}
  \textbf{\bibinfo{volume}{109}}, \bibinfo{pages}{1492--1505},
  \doiprefix\url{10.1103/PhysRev.109.1492} (\bibinfo{year}{1958}).

\bibitem{RuanNL2014}
\bibinfo{author}{Wang, Y.} \emph{et~al.}
\newblock \bibinfo{journal}{\bibinfo{title}{Phonon lateral confinement enables
  thermal rectification in asymmetric single-material nanostructures}}.
\newblock {\emph{\JournalTitle{Nano Letters}}} \textbf{\bibinfo{volume}{14}},
  \bibinfo{pages}{592--596}, \doiprefix\url{10.1021/nl403773f}
  (\bibinfo{year}{2014}).

\bibitem{ma2003self}
\bibinfo{author}{Ma, Y.}, \bibinfo{author}{Yu, B.}, \bibinfo{author}{Zhang, D.}
  \& \bibinfo{author}{Zou, M.}
\newblock \bibinfo{journal}{\bibinfo{title}{A self-similarity model for
  effective thermal conductivity of porous media}}.
\newblock {\emph{\JournalTitle{Journal of Physics D: Applied Physics}}}
  \textbf{\bibinfo{volume}{36}}, \bibinfo{pages}{2157} (\bibinfo{year}{2003}).

\bibitem{ma2004fractal}
\bibinfo{author}{Ma, Y.}, \bibinfo{author}{Yu, B.}, \bibinfo{author}{Zhang, D.}
  \& \bibinfo{author}{Zou, M.}
\newblock \bibinfo{journal}{\bibinfo{title}{Fractal geometry model for
  effective thermal conductivity of three-phase porous media}}.
\newblock {\emph{\JournalTitle{Journal of applied physics}}}
  \textbf{\bibinfo{volume}{95}}, \bibinfo{pages}{6426--6434}
  (\bibinfo{year}{2004}).

\bibitem{miller1969bounds}
\bibinfo{author}{Miller, M.~N.}
\newblock \bibinfo{journal}{\bibinfo{title}{Bounds for effective electrical,
  thermal, and magnetic properties of heterogeneous materials}}.
\newblock {\emph{\JournalTitle{Journal of Mathematical Physics}}}
  \textbf{\bibinfo{volume}{10}}, \bibinfo{pages}{1988--2004}
  (\bibinfo{year}{1969}).

\bibitem{archie1942electrical}
\bibinfo{author}{Archie, G.~E.} \emph{et~al.}
\newblock \bibinfo{journal}{\bibinfo{title}{The electrical resistivity log as
  an aid in determining some reservoir characteristics}}.
\newblock {\emph{\JournalTitle{Transactions of the AIME}}}
  \textbf{\bibinfo{volume}{146}}, \bibinfo{pages}{54--62}
  (\bibinfo{year}{1942}).

\bibitem{thovert1990thermal}
\bibinfo{author}{Thovert, J.}, \bibinfo{author}{Wary, F.} \&
  \bibinfo{author}{Adler, P.}
\newblock \bibinfo{journal}{\bibinfo{title}{Thermal conductivity of random
  media and regular fractals}}.
\newblock {\emph{\JournalTitle{Journal of applied physics}}}
  \textbf{\bibinfo{volume}{68}}, \bibinfo{pages}{3872--3883}
  (\bibinfo{year}{1990}).

\bibitem{PhilpottJAP2003}
\bibinfo{author}{Cahill, D.~G.} \emph{et~al.}
\newblock \bibinfo{journal}{\bibinfo{title}{Nanoscale thermal transport}}.
\newblock {\emph{\JournalTitle{Journal of Applied Physics}}}
  \textbf{\bibinfo{volume}{93}}, \bibinfo{pages}{793--818},
  \doiprefix\url{10.1063/1.1524305} (\bibinfo{year}{2003}).

\bibitem{XuAPL2011}
\bibinfo{author}{Hao, F.}, \bibinfo{author}{Fang, D.~N.} \&
  \bibinfo{author}{Xu, Z.~P.}
\newblock \bibinfo{journal}{\bibinfo{title}{Mechanical and thermal transport
  properties of graphene with defects}}.
\newblock {\emph{\JournalTitle{Applied Physics Letters}}}
  \textbf{\bibinfo{volume}{99}}, \doiprefix\url{10.1063/1.3615290}
  (\bibinfo{year}{2011}).

\bibitem{SiMolecules2018}
\bibinfo{author}{Kang, Y.} \emph{et~al.}
\newblock \bibinfo{journal}{\bibinfo{title}{Thermal transport of graphene
  sheets with fractal defects}}.
\newblock {\emph{\JournalTitle{Molecules}}} \textbf{\bibinfo{volume}{23}},
  \doiprefix\url{10.3390/molecules23123294} (\bibinfo{year}{2018}).

\bibitem{RanGeSi2018}
\bibinfo{author}{Ran, X.}, \bibinfo{author}{Guo, Y.}, \bibinfo{author}{Hu, Z.}
  \& \bibinfo{author}{Wang, M.}
\newblock \bibinfo{journal}{\bibinfo{title}{Interfacial phonon transport
  through si/ge multilayer film using monte carlo scheme with spectral
  transmissivity}}.
\newblock {\emph{\JournalTitle{Frontiers in Energy Research}}}
  \textbf{\bibinfo{volume}{6}}, \bibinfo{pages}{28},
  \doiprefix\url{10.3389/fenrg.2018.00028} (\bibinfo{year}{2018}).

\bibitem{MajumdarJHT2002}
\bibinfo{author}{Abramson, A.~R.}, \bibinfo{author}{Tien, C.~L.} \&
  \bibinfo{author}{Majumdar, A.}
\newblock \bibinfo{journal}{\bibinfo{title}{Interface and strain effects on the
  thermal conductivity of heterostructures: A molecular dynamics study}}.
\newblock {\emph{\JournalTitle{Journal of Heat Transfer-Transactions of the
  {ASME}}}} \textbf{\bibinfo{volume}{124}}, \bibinfo{pages}{963--970},
  \doiprefix\url{10.1115/1.1495516} (\bibinfo{year}{2002}).

\bibitem{ZurbuchenNMat2014}
\bibinfo{author}{Ravichandran, J.} \emph{et~al.}
\newblock \bibinfo{journal}{\bibinfo{title}{Crossover from incoherent to
  coherent phonon scattering in epitaxial oxide superlattices}}.
\newblock {\emph{\JournalTitle{Nature Materials}}}
  \textbf{\bibinfo{volume}{13}}, \bibinfo{pages}{168--172},
  \doiprefix\url{10.1038/Nmat3826} (\bibinfo{year}{2014}).

\end{thebibliography}

\section*{Acknowledgements}
This work was supported as a part of the Computational Materials Sciences Program funded by the U.S Department of Energy, Office of Science, Basic Energy Sciences, under Award Number DE-SC0014607. All Simulations were performed at the Center for High Performance Computing of the University of Southern California

\section*{Author contributions statement}

P.V, R.K.K and A.N conceived the simulations. A.K and N.B performed simulations and data analysis. All authors wrote and reviewed the manuscript.

\end{document}